\documentclass[reprint,
superscriptaddress,
 amsmath,amssymb,
 notitlepage,
aps,
prx,
10pt, 
tightenlines
]{revtex4-2}

\usepackage[utf8]{inputenc}
\usepackage[caption=false]{subfig}
\usepackage{graphicx}
\usepackage{dcolumn}
\usepackage{bm}
\usepackage{xcolor}
\usepackage{verbatim}
\usepackage{tabularx}
\usepackage{bbold}
\usepackage{float}
\usepackage{bm}
\usepackage{bbm}
\usepackage[colorlinks=True, citecolor=blue, urlcolor=blue, linkcolor=blue]{hyperref}
\usepackage{booktabs}
\usepackage{natbib}

\usepackage{braket}
\usepackage[ruled,vlined]{algorithm2e}
\usepackage{comment}

\definecolor{crimson}{rgb}{.8, 0, 0}

\makeatletter
\newcommand{\showtextwidth}{
    \strip@pt\textwidth
}
\makeatother

\begin{document}

\title{Quantum quench dynamics as a shortcut to adiabaticity}
\author{Alexander~Lukin}
\thanks{These authors contributed equally to this work}
\email[\newline Corresponding author: ]{alukin@quera.com}
\affiliation{QuEra Computing Inc., 1284 Soldiers Field Road, Boston, MA 02135, USA}

\author{Benjamin~F.~Schiffer}
\thanks{These authors contributed equally to this work}
\email[\newline Corresponding author: ]{alukin@quera.com}
\affiliation{Max-Planck-Institut f\"ur Quantenoptik, Hans-Kopfermann-Str.~1, D-85748 Garching, Germany}%

\author{Boris~Braverman}
\affiliation{QuEra Computing Inc., 1284 Soldiers Field Road, Boston, MA 02135, USA}
\affiliation{Department of Physics, University of Toronto, Toronto, ON M5S 1A7, Canada}

\author{Sergio~H.~Cantu}
\author{Florian~Huber}
\author{Alexei~Bylinskii}
\author{Jesse~Amato-Grill}
\affiliation{QuEra Computing Inc., 1284 Soldiers Field Road, Boston, MA 02135, USA}

\author{Nishad~Maskara}
\author{Madelyn~Cain}
\affiliation{Department of Physics, Harvard University, Cambridge, MA 02138, USA}%

\author{Dominik S.~Wild}
\affiliation{Max-Planck-Institut f\"ur Quantenoptik, Hans-Kopfermann-Str.~1, D-85748 Garching, Germany}%

\author{Rhine~Samajdar}
\affiliation{Department of Physics, Princeton University, Princeton, NJ 08544, USA}%
\affiliation{Princeton Center for Theoretical Science, Princeton University, Princeton, NJ 08544, USA}

\author{Mikhail~D.~Lukin}
\affiliation{Department of Physics, Harvard University, Cambridge, MA 02138, USA}%

\date{\today}

\begin{abstract}

 The ability to efficiently prepare ground states of quantum Hamiltonians via adiabatic protocols is typically limited by the smallest energy gap encountered during the quantum evolution. This presents a key obstacle for quantum simulation and realizations of adiabatic quantum algorithms in large systems, particularly when the adiabatic gap vanishes exponentially with system size. Using QuEra's Aquila programmable quantum simulator based on Rydberg atom arrays, we experimentally demonstrate 
 a method to circumvent such limitations. Specifically, we develop and test a ``sweep-quench-sweep'' quantum algorithm in which the incorporation of a quench step serves as a remedy to the diverging adiabatic timescale. These quenches introduce a macroscopic reconfiguration between states separated by an extensively large Hamming distance, akin to quantum many-body scars. 
 Our experiments show that this approach significantly outperforms the adiabatic algorithm, illustrating that such quantum quench algorithms can provide a shortcut to adiabaticity for large-scale many-body quantum systems. 
\end{abstract}

\maketitle

\section{Introduction}
The advent of quantum computers and quantum simulators has opened new avenues for studies of strongly correlated quantum systems~\cite{altman2021quantum}. Central to such endeavors is the efficient preparation of desired 
quantum many-body \textit{ground} states. Besides being of fundamental interest for simulating and understanding complex quantum systems,
the ground states of specific Hamiltonians can also encode the solutions~\cite{Ebadi2022Quantum, Dwave-sg, IsingQC2, IsingQC1} to---and bear deep implications for---useful quantum combinatorial optimization~\cite{Aeppli99,Farhi2001Quantum} and sampling problems~\cite{Somma2007Quantum, Wild2021Quantum}.

\begin{figure}[t!]
    \includegraphics[width=1.0\linewidth]{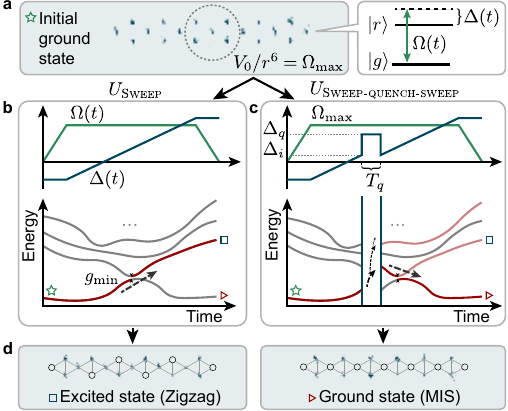}
    \caption{Quantum quench as a shortcut to adiabaticity. \textbf{(a)} Fluorescence image of atoms in their ground state arranged in a quasi-1D geometry. Each atom is simultaneously driven to the Rydberg state $\ket{r}$ with a time-dependent Rabi frequency $\Omega(t)$ and detuning $\Delta(t)$. Pairs of atoms in the Rydberg state interact via a long-ranged van der Waals potential that falls off with interatomic distance as $1/\lvert\textbf{r}\rvert^6$.
    The dashed circle marks the ``blockade radius'', defined by the distance where the interaction energy matches $\Omega_\text{max}$. \textbf{(b)} An adiabatic protocol, implemented by a linear sweep of the detuning, fails to prepare the target ground state due to a diabatic transition to an excited state at a small gap along the path. \textbf{(c)} A quantum quench prior to crossing the small gap transfers a macroscopic state fraction to the first excited state that diabatically connects to the ground state during the subsequent linear sweep. \textbf{(d)} Fluorescence images of the final configuration. Circles represent atoms in the Rydberg state (dark during imaging) and lines the connectivity according to the Rydberg blockade.  Left: one of two configurations of the zigzag state corresponding to the final first-excited state; right: the true ground state with the maximum number of Rydberg excitations.
    }
    \label{fig:Fig1}
\end{figure}

In general, preparing a quantum many-body ground state is a QMA-hard problem in the worst case~\cite{Kempe2006complexity}; hence, no general-purpose algorithm is expected to exist that achieves this task in polynomial runtime on a quantum computer. A commonly used heuristic approach is the quantum adiabatic algorithm (QAA), which relies on transforming an initial easily prepared ground state to a target state by slowly varying some parameters of the Hamiltonian~\cite{albash2018adiabatic}. 
The required runtime for success of the QAA depends inversely on the minimum gap between the ground state and the first excited state~\cite{Amin2009Consistency, Jansen2007Bounds}. 
Since hardware imperfections place stringent limitations on the duration of coherent evolution in quantum simulators, this poses a major challenge for the QAA in the presence of small gaps. 
Here, we present a scalable method to mitigate this problem for a class of systems with a superexponentially small gap that the QAA cannot circumvent.

Due to the constraints on the QAA described above, alternative approaches to ground state preparation
have been extensively explored. These include 
variational methods 
in which the quantum circuit or dynamics are parametrized and then optimized in a quantum-classical feedback loop with a suitable cost function~\cite{Farhi2014quantum,cerezo2021variational, Bharti2022Noisy}. However, their efficiency can be severely limited by prohibitive training costs due to many parameters and (noise-induced) barren plateaus~\cite{Wang2020Noise, Larocca2024Review}.
Another set of approaches include \emph{shortcuts to adiabaticity} (STA)~\cite{torrontegui2013shortcuts,Werschnik2007Quantum,GueryOdelin2019Shortcuts}, which, however, are challenging to generalize to large systems~\cite{xu2020effects} due to the absence of precise knowledge of the complex many-body wavefunction~\cite{saberi2014adiabatic} or the need for nonlocal multi-body terms in the driving Hamiltonians that are difficult to realize in practical quantum simulators~\cite{del2013shortcuts}.

In this paper, inspired by the suggestion of aiding the QAA by starting in an excited state~\cite{crosson2014different}, we propose and experimentally test an approach 
combining quasi-adiabatic sweeps with a quantum quench
as a shortcut to adiabaticity. For a system where the adiabatic gap is prohibitively small, our approach dramatically improves the target-state fidelity, while optimizing over only three parameters. Specifically, we focus on Ising chains with long-ranged interactions and constrained dynamics (extending the so-called PXP model~\cite{Bernien2017Probing}) arranged in particular quasi-1D and 2D geometries [see, e.g., Fig.~\ref{fig:Fig1}(a)]. In preparing the many-body ground states---which represent solutions to the Maximum Independent Set (MIS) problem on the associated graph---the QAA breaks down owing to the presence of a superexponentially small gap at a first-order transition, as illustrated in Fig.~\ref{fig:Fig1}(b). We remedy this by using a variationally optimized \emph{sweep-quench-sweep} (SQS) protocol, depicted schematically in Fig.~\ref{fig:Fig1}(c): the quantum quench during the quasiadiabatic sweep induces a global reconfiguration of the state, akin to many-body scar dynamics~\cite{Bernien2017Probing}. Experimentally testing this protocol using arrays of up to 73 neutral atom qubits,
we find orders-of-magnitude improvements compared to the QAA in the preparation of target [Fig.~\ref{fig:Fig1}(d)] ground states. These results demonstrate the utility of quantum quenches as an algorithmic tool for the extension of STA methods to strongly interacting many-body systems.

\begin{figure*}[t!]
    \includegraphics[width=1.0\linewidth]{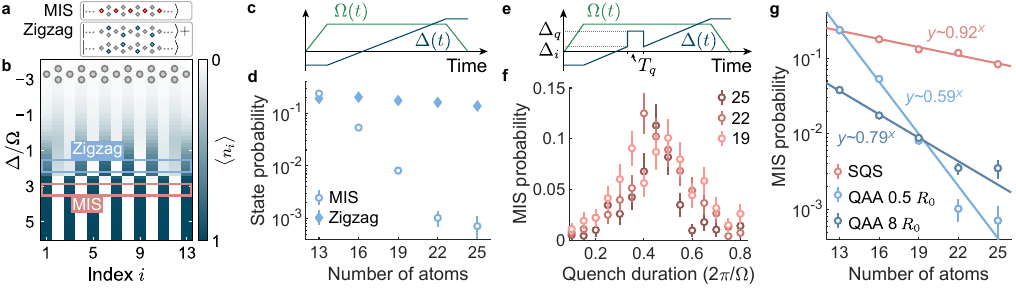}
    \caption{\textbf{(a)} MIS and zigzag configurations (schematic) of the quasi-1D chain. For a given chain length, the zigzag state has one fewer excitation than the MIS (Rydberg excitations are highlighted in color). 
    \textbf{(b)} The density plot shows the local Rydberg occupation $\langle n_i \rangle$ (with doublets summed together) of the many-body ground state at different $\Delta/\Omega$. The competition between the Rydberg excitation energy and the total interaction cost of the configuration results in a zigzag-like ground state for small positive detunings and the MIS for large detunings, separated by a first-order transition. 
    \textbf{(c)} The QAA is realized by sweeping the detuning at a constant rate of $\dot{\Delta}$\,$=$\,$\partial \Delta/\partial t$\,$=$\,$0.5\;R_0$, where $R_0=\Omega^2/(2\pi)$, at a constant Rabi frequency. \textbf{(d)} After a linear sweep, the probability of finding the MIS state decreases much more rapidly than that of the zigzag state as the system size increases. This indicates the breakdown of the QAA. \textbf{(e)} During a linear sweep, the SQS protocol introduces a discontinuous jump of the detuning from $\Delta_i$ to $\Delta_q$ for the duration $T_q$; 
    then, the linear sweep continues. \textbf{(f)} For quench parameters $\Delta_i=0.55\;\Omega, \Delta_q= 1.5\;\Omega$, and $\dot{\Delta} = 1.75\;R_0$, the MIS probability peaks at a certain duration of the quench with the magnitude enhanced nearly hundredfold compared to a linear sweep for the longest chain probed. In addition, the optimal quench duration is approximately independent of the system size
    (different colors). 
    \textbf{(g)} Comparison of the scaling behavior of the SQS protocol and linear sweeps at different rates. The improved performance of the faster linear sweeps is attributed to diabatic excitations before the first-order transition. The SQS significantly outperforms linear sweeps independent of the rate (see App.~\ref{app:implementation} for details of experimental parameters). Error bars are the standard error of the mean, and for some points, are smaller than the marker size.
    }
    \label{fig:Fig2}
\end{figure*}

\section{Experimental system}
Our setup consists of an array of neutral atoms in optical tweezers; a global laser field drives transitions between the atomic ground state $\ket{g}$ and a highly excited Rydberg state $\ket{r}$ for each atom simultaneously. The dynamics of the system are described by the many-body Hamiltonian
\begin{align}
\label{eq:HRyd}
   \frac{H_\text{Ryd}}{\hbar} = \frac{\Omega}{2} \sum_{i } \sigma_i^x  - \Delta \sum_{i } n^{}_i 
    + V^{}_0&\sum_{i, j} \frac{n^{}_i n^{}_j}{|\textbf{r}^{}_i-\textbf{r}^{}_j|^6},
\end{align}
where $\Omega$ denotes the ground-Rydberg Rabi frequency with $\sigma_i^x$\,$\equiv$\,$ (\ket{g}\bra{r}+\ket{r}\bra{g})_i$, $\Delta$ is the laser detuning, and $V_0$ parametrizes the strength of the van der Waals interactions between two atoms $i,j$ at positions $\textbf{r}^{}_i, \textbf{r}^{}_j$ in the Rydberg state ($n_i$\,$\equiv$\,$\ket{r}_i\bra{r}$); see Fig.~\ref{fig:Fig1}(a) and Appendix~\ref{app:implementation}. We experimentally implement this Hamiltonian on QuEra Computing's Aquila device~\cite{Wurtz2023Aquila} and systematically probe its coherent quantum dynamics.

At large \emph{negative} detunings $\Delta$, the many-body ground state of the system has all atoms in their individual atomic ground state $\ket{g}$, as any Rydberg excitation increases the total energy of the system. When $\Delta$ is large and \emph{positive}, the system seeks to maximize the number of atoms in the Rydberg state $\ket{r}$, subject to the occupation constraints imposed by the long-ranged interaction potential ${V^{}_0}/{|\textbf{r}^{}_i-\textbf{r}^{}_j|^6}$. This allows one to choose the distance between the atoms such that for nearest neighbors (NN), the interaction strength $V^{}_{nn} \gg \Omega$, whereas for beyond nearest neighbors, $V^{}_{bnn} < \Omega$. In this case, the system is commonly approximated by the so-called PXP model~\cite{Lesanovsky2012Interacting} where the simultaneous excitations of neighboring atoms are prohibited. This approach has been used to encode the solutions of the unit-disk graph MIS problem onto the ground states of Rydberg atom arrays ~\cite{Pichler2018Quantuma, Ebadi2022Quantum,cain2023quantum,PRXQuantum.5.020313}.

\section{Results}
\subsection{Adiabatic protocol breakdown}
We first focus on a geometric arrangement where a single atom alternates with a doublet of atoms, forming a chain of equilateral triangles that we term a \emph{Rydberg doublet chain}~[Fig.~\ref{fig:Fig1}(a)]. We consider a side length of $5.5$~\textmu m resulting in $V_{nn}$\,$=$\,$12.5\;\Omega_\text{max}$ and $V_{bnn}$\,$<$\,$0.4\;\Omega_\text{max}$. In the PXP approximation, the many-body ground state for positive detunings corresponds to the Rydberg excitations arranged along the middle row, thereby maximizing the number of atoms in $\vert r\rangle$ subject to the NN constraint [Fig.~\ref{fig:Fig1}(d), right]. We refer to this as the MIS state, in analogy with the graph problem of the same name. 

Experimentally, we first attempt to prepare the many-body ground state using the QAA by performing a slow linear sweep of the detuning. The resulting probability of finding the MIS state at the end of the protocol, $P_{\textsc{mis}}$, exhibits a fast exponential decrease with the system size [Fig.~\ref{fig:Fig2}(d)]. Importantly, for the long chains, the MIS state is not the most likely outcome after the adiabatic sweep. Instead, there are two states (occurring with roughly equal probability) where the Rydberg excitations are ordered in a zigzag-like pattern in the middle of the chain, resulting in one fewer excitation compared to the MIS state [Fig.~\ref{fig:Fig2}(a)]. In addition, the probability of finding one of these zigzag states at the end of the protocol decreases much more slowly with the system size than the MIS [Fig.~\ref{fig:Fig2}(d)].

To explain this experimental observation we need to go beyond the PXP approximation. The soft-core nature of the Rydberg potential results in two distinct next-nearest-neighbor (NNN) interactions: $V_h$ ($V_d$) for atoms situated horizontally (diagonally) apart [Fig.~\ref{fig:App_Multiplex}(b,c)]. For the chosen geometry, $V_h$\,$\approx$\,$2.37\;V_d$, so the competition between these NNN interactions and the energy difference of exciting an additional Rydberg atom can result in the observed zigzag configuration being the true many-body ground state for small positive detunings [Fig.~\ref{fig:Fig2}(b)]. At a critical detuning $(\Delta/\Omega)_\text{crit}$, a first-order transition occurs between the zigzag and MIS states [Fig.~\ref{fig:Fig2}(b)], with a spectral gap that scales superexponentially as $\exp(-N \log N)$ with the system size $N$. This \emph{perturbative crossing}~\cite{Amin2009First} arises because the system, during the adiabatic algorithm, gets driven into a configuration which is very far in Hamming distance from the target state.
The dynamics therefore stay quasi-adiabatic until this small gap is encountered, whereupon Landau-Zener transitions diabatically transfer population to the first excited state, thereby reducing $P_{\textsc{mis}}$.

\begin{figure}[tb]
    \includegraphics[width=1.0\linewidth]{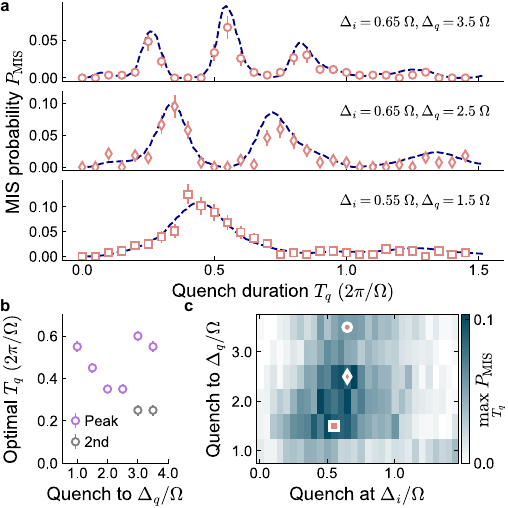}
    \caption{Stability analysis of the SQS protocol for a chain of 22 atoms. \textbf{(a)} Representative traces of the MIS probability as a function of the quench duration $T_q$ for different quench parameters $\Delta_i$ and $\Delta_q$. The time to the first MIS revival decreases with increasing $\Delta_q$; for higher final detunings, multiple revivals are observed. The simulated data (dashed lines) is multiplied by $0.92^8$ to account for misdetection of any of the 8 Rydberg atoms in the MIS~\cite{Wurtz2023Aquila}; a time shift of $0.02\times2\pi/\Omega=8\;$ns is attributed to delays in the system response (App.~\ref{app:implementation}).
    \textbf{(b)} Optimal quench duration $T_q$, corresponding to the maximum MIS probability over the time trace, for different $\Delta_q$, exhibiting an approximately linear dependence. The jump at ${\Delta_q}/{\Omega} = 3$ corresponds to the probability maximum moving from the first revival to the second. \textbf{(c)} The optimal initial detuning is determined by scanning $\Delta_i$ at a fixed $\Delta_q$ (with the corresponding optimal $T_q$). A broad optimal region for $\{\Delta_i, \Delta_q\}$ is apparent. The colored dots indicate the respective time series in (a). Error bars in (a) are the standard error of the mean, and for some points, are smaller than the marker size. Error bars in (b) correspond to the spacing of points in the time traces.}
    \label{fig:Fig3}
\end{figure}

\subsection{Sweep-quench-sweep protocol}
An improvement over the QAA was recently proposed in Ref.~\onlinecite{Schiffer2023Circumventing}, which theoretically showed that quenches from an excited state can introduce revivals in $P_{\textsc{mis}}$, reminiscent of quantum many-body scars~\cite{Bernien2017Probing, Bluvstein2021Controlling}. Motivated by these insights, instead of a single quasiadiabatic sweep, we now introduce a quench between two segments of a continuous ramp, where the detuning is instantaneously changed from a value of $\Delta_i$ to $\Delta_q$ for a quench duration of $T_q$ [Fig.~\ref{fig:Fig2}(e)].

Deploying this protocol for the Rydberg doublet chain we find that the ground-state probability $P_{\textsc{mis}}$ can be dramatically improved compared to the QAA. We first notice that the optimal choice of the quench duration $T_q\approx 0.45/\Omega$ is approximately independent of the chain length [Fig.~\ref{fig:Fig2}(f)]. Using this optimal $T_q$ for a chain of 25 atoms, we are able to boost the ground-state probability by over two orders of magnitude, as seen in Fig.~\ref{fig:Fig2}(g).
We then systematically compare the SQS protocol to linear ramps with different speeds; two such sweeps are shown in Fig.~\ref{fig:Fig2}(g), fitting the MIS probability in each case as $P_{\textsc{mis}}$\,$=$\,$p\, b^N$. For smaller system sizes, the slow sweep outperforms the fast one (since the minimal gap is large enough to allow for maintaining adiabaticity) but for sufficiently long chains, the latter yields a higher $P_{\textsc{mis}}$. Remarkably, we find that the SQS protocol demonstrates superior scaling compared to \textit{all} linear sweeps. Intuitively, this performance enhancement is due to coherent population transfer to the excited states~\cite{muthukrishnan2016tunneling,zhou_quantum_2020} \textit{before} the first-order transition, which is then diabatically transferred back to the ground state at the perturbative crossing due to the small gap~\cite{crosson2014different}. This is also why we see the fast linear ramps yielding an improvement in the scaling of $P_{\textsc{mis}}$ with $N$ over slow sweeps (this, of course, comes at the expense of a reduced final ground-state probability for small system sizes since the total number of populated eigenstates also increases with the ramp speed).

\begin{figure}[b]
    \includegraphics[width=1.0\linewidth]{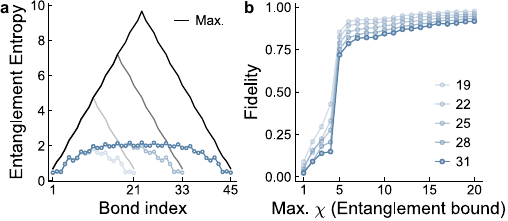}
    \caption{\textbf{(a)} The von Neumann entanglement entropy at the end of the SQS protocol, indicating the absence of volume-law entanglement generation. The dynamics are simulated using tensor-network methods for the Rydberg Hamiltonian ($N=22,34,46$) including up to NNN interactions and allowing for arbitrary bond dimension at a fixed precision threshold $\epsilon$\,$=$\,$10^{-8}$. The upper bounds (drawn in shades of black) are derived using a hard-core interaction potential (PXP model) for a more realistic estimate of the effective Hilbert space dimension.
    \textbf{(b)} 
    We simulate the SQS protocol for various small bond dimensions $\chi$, effectively 
    restricting the dynamics to a Hilbert space with bounded entanglement.
    As a measure of the fidelity of the simulation, we show the norm of the matrix product state following truncation to a fixed bond dimension.
    The results indicate that $\chi=5$ is sufficient to capture the dynamics with high fidelity.
    System-size independent parameters ($T_q = 0.45 \times 2\pi/\Omega, \Delta_i =0.55\;\Omega$, and $\Delta_q= 1.5\;\Omega$) are used for both panels.}
    \label{fig:Fig4}
\end{figure}

\subsection{Scar-like quantum dynamics}
Having observed that the SQS protocol significantly improves the probability for preparing the ground state of the Rydberg doublet chain, we can gain insights into its mechanism by examining the dynamics \textit{during} the quench. This is showcased in Fig.~\ref{fig:Fig3}(a), revealing pronounced oscillations of $P_\textsc{mis}$ as a function of $T_q$, especially for quenches with large $\Delta_q$. These dynamics are phenomenologically similar to quantum many-body scars in which a quench triggers persistent oscillations between two quantum states~\cite{Bernien2017Probing}, but unlike scars, the observed oscillations decay after a few cycles.

In our experiments, we observe that the revival times depend on the final detuning $\Delta_q$ but are independent of the initial quench point $\Delta_i$. For each $\Delta_q/\Omega$, we determine the quench duration $T_q$ that optimizes the probability $P_\textsc{mis}$ [Fig.~\ref{fig:Fig3}(b)]; interestingly, we find that the time of the first revival (which can be regarded as the period of the oscillations) depends near-linearly on the quench detuning. We then fix $\Delta_q$ as well as the corresponding optimal $T_q$ and scan the initial detuning $\Delta_i$ to study the generality of the protocol. Our data reveals that the physics underlying the SQS protocol in the Rydberg doublet chain is robust against variations in quench parameters---as seen in Fig.~\ref{fig:Fig3}(c)---and is \textit{not} a consequence of a fine-tuned point in parameter space.

\begin{figure}[t]
    \includegraphics[width=1.0\linewidth]{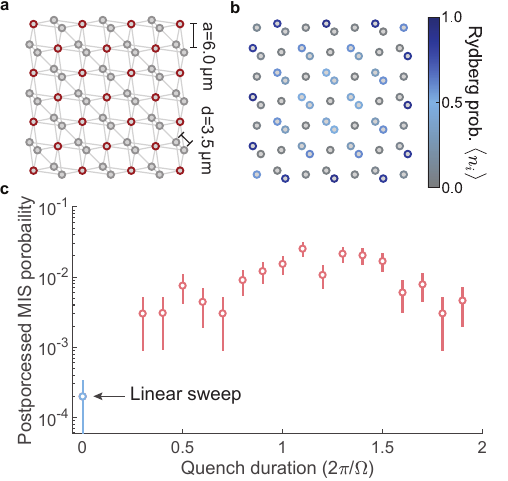}
    \caption{State preparation in two dimensions. \textbf{(a)} Schematic 2D geometry with $N=73$ atoms; the gray lines illustrate the connectivity graph corresponding to the Rydberg blockade. Excited atoms in the ground state at a large positive detuning are highlighted in red (MIS state). \textbf{(b)}~Measured Rydberg probabilities after a linear sweep, with $\dot{\Delta} = 1.5\;R_0$, showing a high probability of preparing the 2D analog of the zigzag-like configuration observed in the quasi-1D geometry, which has a large Hamming distance to the MIS state. \textbf{(c)}~For quench parameters close to those optimal for the quasi-1D chain ($\Delta_i$\,$=$\,$0.5\;\Omega, \Delta_q= 1.5\;\Omega$, and $\dot{\Delta} = 1.5\;R_0$), a peak of $P_{\textsc{mis}}$ is observed as a function of the quench duration. The data point at zero quench duration corresponds to a simple linear sweep. The SQS sequence boosts the MIS probability by two orders of magnitude compared to the linear sweep. A postprocessing algorithm mitigates imperfections in the Rydberg detection process as detailed in App.~\ref{app:implementation}. Error bars are the standard error of the mean.}
    \label{fig:Fig5}
\end{figure}

To understand the origin of the observed scar-like dynamics, we turn to numerical tensor-network calculations.
Typically, scarred dynamics arise from a small set of low-entanglement excited states which support the oscillating dynamics~\cite{Turner2018Weak,ho2019periodic,Serbyn2021Quantum,chandran2023quantum}. 
Here, we find that the dynamics of the SQS protocol are also scar-like in the sense that they primarily occur within a low-entanglement subspace. 
First, considering the entanglement entropy at a central bipartition of the chain, we find that the SQS procedure does not generate volume-law entanglement~[Fig.~\ref{fig:Fig4}(a)], as would have been typical for general chaotic quench dynamics~\cite{mitra2018quantum}.
Further, we repeat our tensor-network simulations while restricting the maximum bond dimension of the matrix product state ansatz (which relates to an upper bound on the entanglement entropy across any bipartition of the system). We observe that the dominant features of the dynamics are preserved for small $\chi$\,$=$\,$ 5$, and this property persists for larger system sizes~[Fig.~\ref{fig:Fig4}(b)].
Since this rather small bond dimension captures the essential features of the time evolution, it indicates that the post-quench dynamics occur in a relatively small corner of the Hilbert space.
We further observe that the quench preferentially excites a few low-lying eigenstates, associated with the MIS and zigzag ordering~(Fig.~\ref{fig:AppSpectra}).
Taken together, these observations suggest that the mechanism which enables the SQS protocol is the ability of the quench to mix the various low-energy competing orders, while only weakly coupling to thermalizing states.

\subsection{Extension to two dimensions}
To demonstrate that the SQS protocol is not restricted to quasi-1D systems, we consider a large class of graphs, which are defined by the existence of cliques that facilitate local reconfigurations of Rydberg excitations to generate exponentially many first-excited states above a \textit{unique} ground state. Such local degeneracies are the key ingredients for graphs in this family, across dimensions, and result in the adiabatic gap decreasing superexponentially with system size. 
One example of such a graph,  illustrated in Fig.~\ref{fig:Fig5}(a)~\cite{Schiffer2023Circumventing}, is constructed from a square grid of atoms, spaced at $a$=6.5\;\textmu m, by replacing every second atom with a diagonally oriented doublet in which atoms are $d$=3\;\textmu m apart. In Fig.~\ref{fig:Fig5}(b), we show that for such a graph with $N$\,$=$\,$73$ atoms, a standard adiabatic sweep yields a state with a suboptimal configuration of Rydberg excitations, akin to the zigzag-type first excited state of the target Hamiltonian in the quasi-1D case. However, using the SQS protocol, we observe in Fig.~\ref{fig:Fig5}(c) that the probability to prepare the MIS configuration can once again be significantly boosted, by two orders of magnitude.

\section{Discussion and outlook}
Our experiments demonstrate that the SQS protocol can serve as a new algorithmic tool for ground state preparation. It is important to note that our approach is 
based on quantum dynamics that are natively available without the overhead typically associated, e.g., with 
Trotter decomposition. 
By diabatically populating the low-lying excited states before a first-order transition~\cite{crosson2014different}---as suggested in the context of quantum annealing~\cite{muthukrishnan2016tunneling} and variational QAOA~\cite{zhou_quantum_2020} methods earlier---this protocol directly implements a shortcut to adiabaticity, which is similar in spirit to so-called bang-bang techniques~\cite{alonso2016generation,cohn2018bang} within the broader landscape of STA methods. 

To place our work more generally in the context of quantum algorithms, the sweep-quench-sweep protocol offers an intriguing new kind of algorithmic framework which uses hardware-efficient dynamics and is complementary to both the
(variational) quantum adiabatic algorithm (VQAA)~\cite{Farhi2000Quantum, Schiffer_2022} and the quantum approximate optimization algorithm (QAOA)~\cite{Farhi2014quantum}. Besides the performance improvements of the SQS protocol over the QAA demonstrated by our experiments, another advantage of our approach is that only few parameters are required in the optimization process.

A common feature of the class of graphs that we study here is their highly regular structure. These particular lattice geometries lead to the prohibitive scaling of the adiabatic gap, but at the same time, they give rise to scar-like dynamics that enable 
the success of the SQS protocol. For random lattices, such as the site-diluted king's graphs~\cite{Samajdar2020Complex,Ebadi2020Quantum,Kalinowski2022Bulk} studied in Ref.~\onlinecite{Ebadi2022Quantum}, quench dynamics could help avoid potential algorithmic slowdowns arising in a clique of the graph. Hence, it would be interesting to explore the possible advantages of the SQS protocol for random graphs, and its generalizability to such problems remains an open question for future investigation.

\begin{acknowledgments}
B.S.~and D.S.W.~thank S.~Hassinger, E.~Kessler, P.~Komar, and the AWS team for computational time on Aquila in the early stage of the project. This work was supported by the Defense Advanced Research Projects Agency (DARPA ONISQ, grant no.~W911NF2010021). Work at Harvard was additionally supported by the US Department of Energy (DOE Quantum Systems Accelerator Center (contract no.~7568717)), the National Science Foundation, the Center for Ultracold Atoms, the NSF Physics Frontiers Center, and the Department of Defense Multidisciplinary University Research Initiative (ARO MURI, grant no.~W911NF2010082).
Work in Munich was part of the Munich Quantum Valley, which is supported by the Bavarian state government with funds from the Hightech Agenda Bayern Plus.
N.M.~acknowledges support by the Department of Energy Computational Science Graduate Fellowship under award number DE-SC0021110. R.S.~is supported by the Princeton Quantum Initiative Fellowship.
\end{acknowledgments}

\appendix

\section{Details of the experimental implementation} \label{app:implementation}

\begin{figure}[b]
    \includegraphics[width=1.0\linewidth]{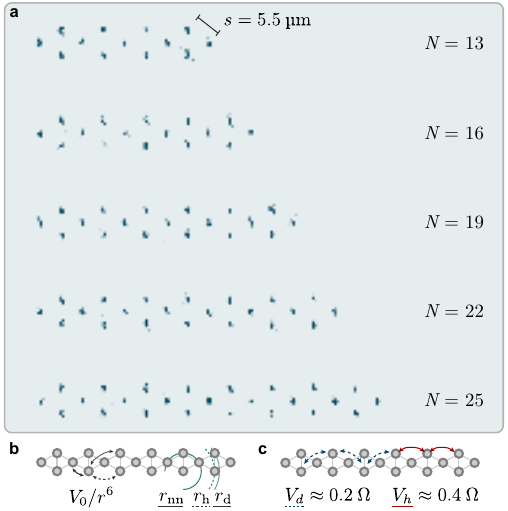}
    \caption{\textbf{(a)} Fluorescence image of the atoms in their ground state for several different total numbers of atoms $N$. We simultaneously load the geometries shown to reduce the total number of experimental cycles.
    \textbf{(b)} Schematic illustration of the relevant interaction energies for the Rydberg doublet chain. The nearest-neighbor distances $r_\text{nn}$ are the same throughout for a geometry of equilateral triangles. Diagonally situated atoms, at a distance $r_\text{d}$, are slightly further apart than horizontally placed atoms, which have a spacing of $r_\text{h}$. \textbf{(c)} The interaction strength decreases as a power law with the distance, giving rise to a horizontal interaction $V_h$ that is approximately double the diagonal interaction $V_d$.}
    \label{fig:App_Multiplex}
\end{figure}

The experimental platform is a programmable array of neutral atoms in optical tweezers called \textit{Aquila} realized by QuEra Computing Inc. A whitepaper with technical specifications is available~\cite{Wurtz2023Aquila}. In this experiment, the technical constraints are somewhat relaxed compared to the cloud-accessible parameters described in Ref.~\onlinecite{Wurtz2023Aquila}. A single experimental cycle begins by laser cooling a cloud of several thousand $^{87}\textrm{Rb}$ atoms, followed by loading the atoms into a reconfigurable two-dimensional geometry defined by a holographic pattern projected using a spatial light modulator. The atoms are quasideterministically sorted into the desired target configuration and optically pumped into the initial ground state. Excitation to Rydberg states is driven by a two-photon transition using two counterpropagating lasers at $420\;\mathrm{nm}$ and $1013\;\mathrm{nm}$, $1\;\mathrm{GHz}$ detuned from the $\mathrm{{}^6P_{3/2}}$ intermediate state.

In order to minimize the impact of systematic fluctuations in the experimental apparatus, for the quasi-1D case, we simultaneously load patterns with $N=13,16,19,22,25$ atoms in a single two-dimensional geometry, shown in Fig.~\ref{fig:App_Multiplex}(a). The typical Rabi frequency used in these experiments was $\Omega = 2.5\;\mathrm{MHz}$. For each chain within the experimental run, the data is postselected on having all atoms in the ground state before the quantum evolution. We have a typical perfect loading fraction of $\approx 80\% , 89\% , 88\% , 90\% ,$ and $71\%$ for a chain length of 13, 16, 19, 22, and 25 atoms respectively, given by the sorting strategy and details of the geometry.

Due to the imperfections in the detection process, the probability of misdetecting a Rydberg atom as a ground-state atom is 8\%, while the probability of misdetecting a ground-state atom as a Rydberg atom is 1\%. In Fig.~\ref{fig:Fig3}(a), as a first-order correction to the noiseless numerical simulation, we therefore rescale the obtained values for $P_\textsc{MIS}$ by $0.92^8$ because the MIS state for a Rydberg doublet chain of 15 atoms has 8 Rydberg atoms. Additionally, we note that the sweep-quench-sweep protocol has a rectangular shape for the time-dependent detuning. This would require an infinitely fast change of the detuning frequency. In the experiment, the electro-optic components have a finite rise and fall time, which effectively leads to a delay in the system response that we fit by hand to a value of 8 ns in Fig.~\ref{fig:Fig3}(a).

A single realization of the 2D geometry is implemented in an individual experimental run. Similarly to the quasi-one-dimensional case, the data is initially postselected on having all atoms in the ground state before evolution.

\section{Postprocessing of the experimental data}

The imperfections in the detection process severely limit the ability to detect the perfect MIS ground state for large systems. We mitigate these imaging imperfections by postprocessing the raw data for the 2D arrays only; for the 1D chains, no postprocessing is performed.

Here, we describe the postprocessing algorithm employed. The underlying principle is that we only seek to mitigate imperfect detection, without solving the MIS problem classically. Our postprocessing routine is described by the following pseudocode: \smallskip

\begin{algorithm}[H]
    \begin{enumerate}
        \item Check if all the atoms are present before the quantum dynamics begin.
        \item At the end of the evolution, check if there are blockade violations.
        \item Flip the atom with the most violations from Rydberg to ground (if there are multiple, pick one randomly).
        \item Repeat steps (2) and  (3) until there are no more blockade violations left.
        \item Check if there are atoms that are not blockaded by any neighbors.
        \item Pick a random atom from (5) and flip it from ground to Rydberg.
        \item Repeat steps (5) and (6) until there are no more atoms left in the ground state that are not covered by the blockade.
        \item Use the resulting configuration as an outcome of the evolution.
    \end{enumerate}
  
    \caption{Postprocessing algorithm for 2D arrays}
    \label{alg1}
\end{algorithm}

\begin{figure*}[tb]
    \includegraphics[width=1.0\linewidth]{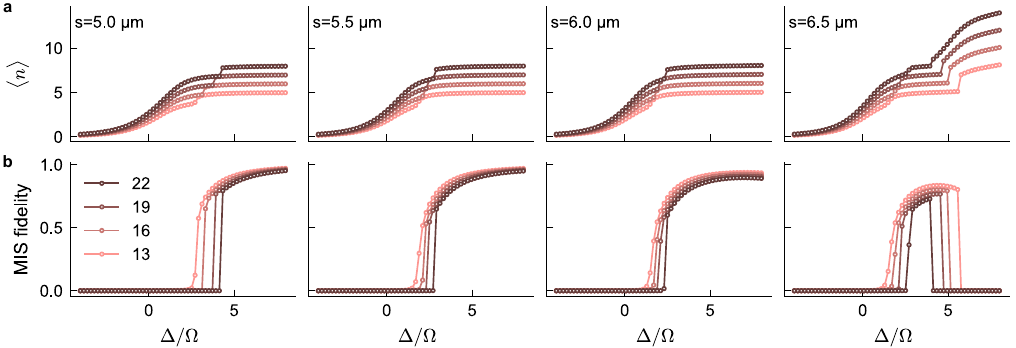}
    \caption{\textbf{(a)} Expected Rydberg excitation number for the ground state of the Rydberg doublet chain at different detunings $\Delta/\Omega$ for different total number of atoms. For a spacing of $s = 6.5$\;\textmu m and large positive detunings, a transition into a blockade-violating state can be observed. \textbf{(b)} The MIS probability of the same ground states. From both observables, it can be seen that the transition to the MIS state occurs for progressively larger $\Delta$ as the chain length is increased.}
    \label{fig:Fig_App_Transition}
\end{figure*}

\section{Geometry of the Rydberg doublet chain}
\subsection{Competing interactions} 
The Rydberg Hamiltonian, as described in the main text,
\begin{align}
   \frac{H_\text{Ryd}}{\hbar} = \frac{\Omega}{2} \sum_{i } \sigma_i^x   - \Delta \sum_{i } n^{}_i 
    + &\sum_{i, j} \frac{V^{}_0 n^{}_i n^{}_j}{|\textbf{r}^{}_i-\textbf{r}^{}_j|^6},
\end{align}
features a long-ranged van der Waals (vdW) interaction between any two atoms in the Rydberg state. The geometry of the Rydberg doublet chain and the interactions between the atoms give rise to a first-order phase transition from the trivial phase to the ordered state. While the interaction between all nearest neighbors is the same for a geometry of equilateral triangles, the interactions between next-nearest neighbors is more subtle. 

In Fig.~\ref{fig:App_Multiplex}(b,c), we highlight the distances between nearest and next-nearest neighbors: diagonally spaced next-nearest neighbors are further apart than horizontally spaced atoms. This results in a weaker interaction between diagonally situated atoms, making a zigzag configuration energetically more favorable than other classical configurations that have one Rydberg excitation less than the MIS.

\subsection{Position of the first-order transition}
Here, we consider the first-order transition to the MIS state for different lattice spacings $s$ in an equilateral geometry in order to identify suitable experimental regimes. This transition occurs at a critical detuning $(\Delta/\Omega)_{\text{crit}}$ and depends on the spacing. We perform DMRG~\cite{Schollwoeck2011density} calculations for the Rydberg doublet chain using the ITensor package~\cite{Fishman_2022}. The ground-state wavefunction of the quasi-1D geometry is represented as a matrix product state (MPS) by snaking through the chain. Our numerics include the interactions between nearest neighbors and next-nearest neighbors, which are necessary to capture the competition between the different MIS-1 states. In Fig.~\ref{fig:Fig_App_Transition}, we show the expectation value of the total Rydberg excitation number $\langle n \rangle = \sum_{i \in \mathcal{V}} \langle n_i \rangle$, and the MIS probability of the respective states. We observe that, as expected, the critical detuning grows with the system size. When the spacing becomes sufficiently large, as seen for $s=6.5$\;\textmu m, blockade violations are possible for large $\Delta/\Omega$. Hence, in the experiment, we set the spacing between the atoms to be $s=5.5$\;\textmu m for the Rydberg doublet chain.

\section{Quantum and classical energetics}
In Ref.~\onlinecite{Schiffer2023Circumventing}, the Rydberg Hamiltonian is analyzed in the limit where the atoms in each doublet are spaced close together. This allows a mapping of the quasi-1D chain to a reduced 1D chain with an enhanced Rabi frequency on alternating sites. Importantly, in this limit, there exists a purely quantum mechanism, stemming from local degeneracies, that gives rise to a superexponentially vanishing adiabatic gap as the chain length is increased. In this paper, however, we observe that including the long-ranged tails of the vdW interaction gives rise to a first order-transition with a superexponentially small gap; as explained in the main text, this can simply be regarded as a classical energetic effect. Hence, there is both a quantum and a classical mechanism at play that independently lead to very similar behaviors. Here, we analyze the difference between the two mechanisms.

First, we note that for both mechanisms the long-ranged tails favor a Rydberg excitation of the boundary atom in the MIS-1 state. We write the left and right boundary terms for a chain of length $L$ as $\ket{B_l}=\ket{r}_1\ket{gg}_2$ and $\ket{B_r}= \ket{gg}_{L-1}\ket{r}_L$. The other atoms compose the bulk of the system.
In the limit where the reduction into a symmetric and antisymmetric subspace holds for each bulk doublet, because the two sites therein are very close to each other, the Rydberg occupations on the even bulk sites are uncorrelated in the MIS-1 state, which is given by 
\begin{equation*}
    \ket{\mathcal{S}} \equiv \ket{B_l} \left( \prod_{i \, \mathrm{odd}} \rvert g \rangle^{}_i \right) \left(\prod_{i \, \mathrm{even}} \frac{\rvert r g \rangle^{}_i +\rvert g r \rangle^{}_i}{\sqrt{2}} \right) \ket{B_r}.
\end{equation*}
The products run over odd and even bulk indices, respectively. This state is involved in the quantum mechanism.

Due to the tails, however, another distinct MIS-1 configuration emerges. In this case, an excitation on one bulk doublet is perfectly correlated with another in a diagonally adjacent position, giving rise to a superposition of the following two zigzag patterns of Rydberg excitations: 
\begin{alignat*}{1}
    \rvert \mathcal{Z} \rangle &= \ket{B_l} \left( \prod_{i \, \mathrm{odd}} \rvert g \rangle^{}_i \right) \left(\rvert r g \rangle^{}_4 \,\rvert g r \rangle^{}_6 \,\rvert r g \rangle^{}_8 \cdots\right) \ket{B_r} ,\\
    \rvert \overline{\mathcal{Z}} \rangle &= \ket{B_l} \left( \prod_{i \, \mathrm{odd}} \rvert g \rangle^{}_i \right) \left(\rvert gr \rangle^{}_4 \,\rvert rg \rangle^{}_6 \,\rvert gr \rangle^{}_8 \cdots\right) \ket{B_r} .
\end{alignat*}
The symmetry between the two sites in each bulk doublet is spontaneously broken in these configurations. 
Note that the (classical) interaction energy for such correlated MIS-1 states $\rvert \mathcal{Z} \rangle, \rvert \overline{\mathcal{Z}} \rangle$ is lower than that for the symmetric, uncorrelated $\ket{\mathcal{S}}$ state because the vdW tails decay as $1/r^6$ with the interatomic distance $r$, which is clearly larger for diagonally spaced excitations. A graphical representation of the $ \rvert \mathcal{Z} \rangle$ (and $\rvert \overline{\mathcal{Z}} \rangle$) state is shown on the left side of Fig.~\ref{fig:Fig1}(d) and in Fig.~\ref{fig:Fig2}(a).

\subsection{Scaling of the critical point for both mechanisms}
The effect of the interaction tails, while different from the local degeneracies, still leads to a similar superexponential scaling of the adiabatic gap. 
The classical configurations corresponding to both the correlated and the uncorrelated MIS-1 states have a Hamming distance $\sim \mathcal{O}(L)$ to the MIS. Considering vdW tails between next-nearest neighbors, the two types of MIS-1 states have a weaker interaction energy (between adjacent even sites) than the MIS (between adjacent odd sites); the resultant \textit{net} difference in the interaction energy is proportional to the length of the chain. 
For the ground state at any given detuning $\Delta$, the energy contributions from the interactions between sites compete with the onsite terms.
Hence, the first-order transition to the MIS state can be predicted to occur in the classical regime of $\Delta_\text{crit}$\,$ \sim$\,$\mathcal{O}(L)$, implying that for longer chains, the transition happens later in the sweep (note that the scaling of $\Delta_\text{crit}$ was $\mathcal{O}(\sqrt{L})$ in the PXP limit). Therefore, the scaling of the adiabatic gap upon including the vdW tails is given by $g_\text{min}$\,$\sim$\,$\mathcal{O}(\exp(-L\log(L))$, 
which is only slightly faster than that for the quantum mechanism of local degeneracies. 

\begin{figure}[b]
    \includegraphics[width=1.0\linewidth]{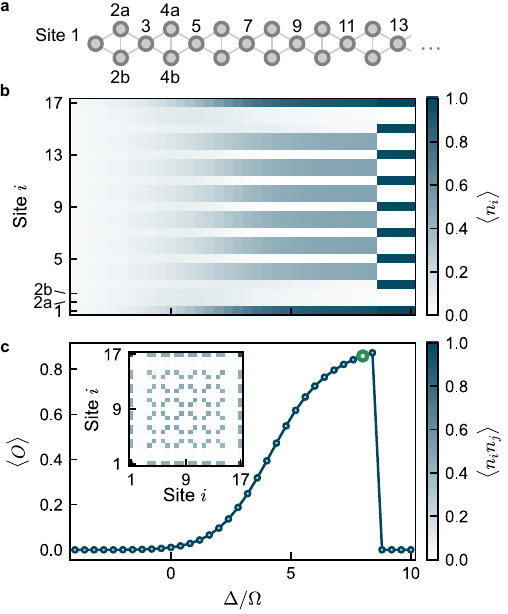}
    \caption{\textbf{(a)} Numbering convention for sites in the Rydberg doublet chain. \textbf{(b)} Local Rydberg occupation of the ground state for a chain of length $L=17$ and a spacing $s=4.5$\;\textmu m, including up to NNN interactions. This tighter spacing shifts the critical point towards larger values of $\Delta$ in order to observe an extended regime of zigzag ordering. For ease of visualizing the zigzag-ordered state, we show $\langle n_i \rangle$ for all atoms individually instead of summing the doublet occupations together. \textbf{(c)} Average order parameter $\langle O \rangle$ and the $\langle n_i n_j\rangle$ correlation matrices of the ground state at $\Delta/\Omega = 8$ (green marker) as an inset. The correlations between Rydberg excitations show clear signatures of zigzag ordering.}
    \label{fig:AppFig_Zigzag}
\end{figure}

\subsection{Transition from quantum to classical mechanism}
Both the quantum and classical mechanisms can be present in actual experimental realizations of the quasi-1D chain. As a function of the detuning, there is a transition from the uncorrelated $\ket{\mathcal{S}}$ state to the correlated $(\ket{\mathcal{Z}}$\,$+$\,$\ket{\overline{\mathcal{Z}}})/\sqrt{2}$ state, and the value of the detuning $\Delta$ at this crossover point is independent of the chain's length for large $L$ since it depends only on the geometry of the chain. The physical properties of the first-order transition to the MIS thus depend on whether the crossover between the different MIS-1 states occurs before or after the first-order transition.
A mean-field model suggests that the crossover between the symmetric and zigzag MIS-1 states occurs at a detuning $\Delta \sim 1/s_y^2$, where $s_y$ is the distance between the doublet sites, while holding the other dimension $s_x$ constant. Consequently, by bringing the doublet sites sufficiently close, the MIS-1 state at the first-order transition into the MIS is expected to be predominantly the uncorrelated $\ket{\mathcal{S}}$ state and the (quantum) local degeneracy mechanism the dominant physical phenomenon. 
Specifically, we can vary the spacing in the horizontal direction $s_x$ for a fixed $s_y$, thus tuning the ratio $s_x/s_y$, which influences the properties of the ground state along an adiabatic sweep via the mechanism described above. 

In order to quantify the correlations between adjacent doublets, we examine a connected two-point correlation function defined as
\begin{align}
\langle O \rangle= \frac{4}{L-7}\sum_{i, j \,\mathrm{even}, \,\alpha} \langle n^{}_{i,\alpha} n^{}_{j,\bar{\alpha}} \rangle -  \langle n^{}_{i,\alpha} \rangle \langle n^{}_{j,\bar{\alpha}} \rangle,
\end{align}
where we average over the $L$\,$-$\,$7$ NNN diagonal pairs between even sites in the bulk. Here, $i,j$ index the position of the doublet along the chain while $\alpha$ labels the top or bottom site within a doublet. In the classical regime ($\Omega /\Delta \rightarrow 0$), this order parameter $\langle O \rangle$ is one for the correlated MIS-1 state and zero for the symmetric MIS-1 or MIS states. We plot $\langle O \rangle$ in Fig.~\ref{fig:AppFig_Zigzag}.

\subsection{Reduced models with purely quantum or classical mechanisms}
\begin{figure}[tb]
    \includegraphics[width=1.0\linewidth]{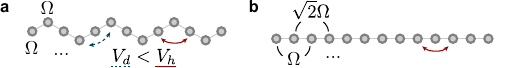}
    \caption{The quantum and classical mechanisms leading to a superexponentially small gap in the Rydberg doublet chain can be seen clearly by considering two simplified geometries that exhibit \textbf{(a)} only a classical mechanism, or \textbf{(b)} a purely quantum mechanism (cf.~Ref.~\onlinecite{Schiffer2023Circumventing}). In the former, the geometry is simply a 1D Rydberg chain perturbed into a zigzag pattern. This leads to different interaction strengths between next-nearest neighbor pairs and a superexponentially small gap.}
    \label{fig:ReduChains}
\end{figure}

Additionally, we briefly introduce two simplified one-dimensional graphs that highlight the difference between the distinct mechanisms leading to a superexponential gap.
The classical one can be mimicked by considering a true 1D chain that is arranged in a zigzag-type pattern [Fig.~\ref{fig:ReduChains}(a)]. Here, the NNN interaction between horizontal atoms $V_h$ is smaller than that between diagonal atoms $V_d$. Hence, as the chain length is increased, the transition from the MIS-1 to the MIS state occurs for ever larger detunings $\Delta/\Omega$.
The quantum mechanism, however, can be encoded in a 1D Rydberg chain where every second (even-numbered) site has an enhanced Rabi frequency [Fig.~\ref{fig:ReduChains}(b)]. An enhancement factor of $k=\sqrt{2}$ corresponds to the Rydberg doublet chain in the limit of the PXP model, as analyzed in Ref.~\onlinecite{Schiffer2023Circumventing}.

\section{Analysis of the sweep-quench-sweep dynamics}

In this section, we present detailed numerical simulations of the quantum dynamics in the sweep-quench-sweep protocol.

\subsection{Spectrum and populated eigenstates}

\begin{figure}[tb]
    \includegraphics[width=1.0\linewidth]{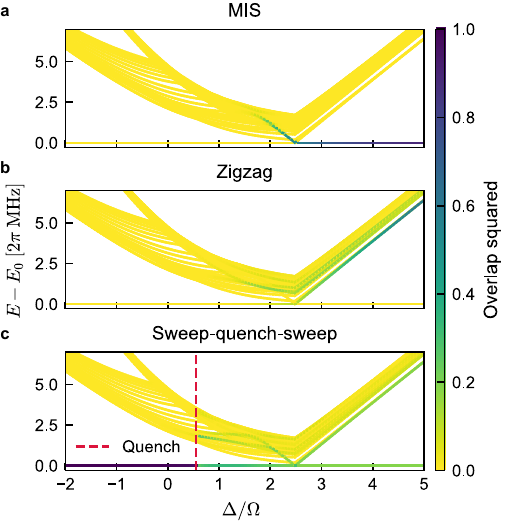}
    \caption{The overlap squared of the lowest eigenstates of $H_\text{Ryd}$ at different values of $\Delta/\Omega$ with \textbf{(a)} the MIS state or \textbf{(b)} the zigzag-type state  are computed for a system of 22 atoms and nearest neighbors separated by $s=5.5$\;\textmu m. In \textbf{(c)}, we show an instance of sweep-quench-sweep dynamics and the instantaneous overlap squared with the eigenstates. The quench shifts population to low-lying excited states which is eventually transferred to the final ground state.}
    \label{fig:AppSpectra}
\end{figure}

We compute the spectra of the Rydberg Hamiltonian for the range of $\Delta / \Omega$ of the quasi-adiabatic sweep using exact diagonalization. Here, we consider the limit where the interaction between nearest neighbors is very large by going to the so-called PXP model~\cite{Lesanovsky2012Interacting}. Additionally, as in the tensor-network simulation, we include finite NNN interactions. The geometry that we consider consists of $N=22$ atoms at a spacing of $s=5.5$\;\textmu m. In good accordance with the DMRG data in Fig.~\ref{fig:Fig_App_Transition}, the position of the smallest gap is around $(\Delta / \Omega)_{\text{crit}}\approx 2.5 $ for this chain length. We color the overlap squared of each low-lying eigenstate with one of two other states of interest: the MIS state [Fig.~\ref{fig:AppSpectra}(a)] and the zigzag-like state $(\ket{\mathcal{Z}}$\,$+$\,$\ket{\overline{\mathcal{Z}}})/\sqrt{2}$ [Fig.~\ref{fig:AppSpectra}(b)]. 

In Fig.~\ref{fig:AppSpectra}(c), we plot the overlap squared of these eigenstates with the instantaneous dynamical state for a sweep-quench-sweep profile that results in a significantly higher final MIS probability than a simple quasi-adiabatic sweep of the same total evolution time. The parameters used for the simulation are $T_q = 0.45 \times 2\pi/\Omega, \Delta_i=0.55\;\Omega$, and $\Delta_q= 1.5\;\Omega$. The position in the sweep where the quench is initialized ($\Delta_i/\Omega$) is indicated with a dotted red line. At this point, a discontinuity of the coloring of the eigenvalues is seen, as a result of the quench dynamics. The quench induces a coherent population transfer to selected low-lying eigenstates, which then continue towards the critical point.

\subsection{Quantum dynamics simulations}

\begin{figure*}[tb]
    \includegraphics[width=1.0\linewidth]{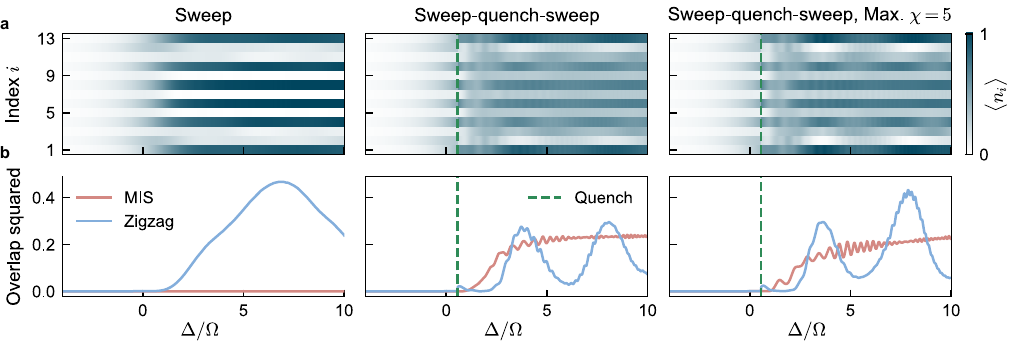}
    \caption{\textbf{(a)} Expected local Rydberg excitation density of the dynamical state as the Hamiltonian is transformed from the initial to the final one. \textbf{(b)} The MIS and zigzag-type state probability of the same dynamical states. The position of the quench is indicated by a dashed, green line. Clearly, the dynamics with and without a quench exhibit very different behaviors. The MIS probability reaches 24\% with a quench. The data in the central column is simulated without setting a maximum bond dimension, while the right column allows for a maximum bond dimension of $\chi=5$ and shows qualitatively similar dynamics.}
    \label{fig:Fig_Dynamics}
\end{figure*}

We also perform simulations of the sweep-quench-sweep dynamics using MPS-based tensor-network methods. The dynamics are implemented using the time-evolving block decimation (TEBD) algorithm~\cite{Vidal2004Efficient} and the truncation cutoff in the singular value decomposition is set to $\epsilon=10^{-8}$. As before, we consider interactions up to next-nearest neighbors and a spacing of $s=5.5$\;\textmu m. The simulated data in Fig.~\ref{fig:Fig3}(a) in the main text is computed with a maximal bond dimension of $\chi=50$ and 1000 discrete time steps in the quasi-adiabatic evolution, which is sufficient for convergence with smaller than the experimental error. In Fig.~\ref{fig:Fig_Dynamics}, we show the expectation value of the local Rydberg excitation density $\langle n_i\rangle$ and the state probability with respect to the MIS and zigzag-type states, respectively. The dynamics are simulated for three different cases. The left panel illustrates a simple sweep, performed at a rate of $1.5\,R_0$. The central panel shows the sweep-quench-sweep dynamics without any bound on the maximum entanglement, with the position of the quench indicated by a green, dashed line. The right panel shows the same sweep-quench-sweep dynamics as the central panel, but restricting the maximum virtual bond dimension of the MPS to $\chi=5$. This rather low value of $\chi$ is the minimal bond dimension that suffices to see a good qualitative agreement with the full dynamics. 

\section{Comparison of linear sweeps at different rates}

\begin{figure}[tb]
    \includegraphics[width=1.0\linewidth]{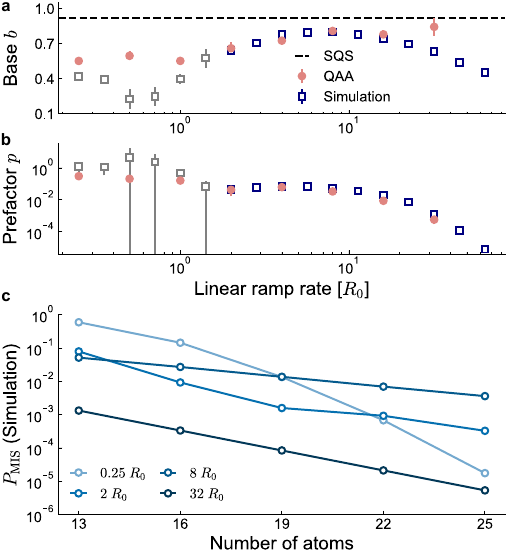}
    \caption{\textbf{(a,b)} Prefactor and base for exponential fits of the data ($P_{\textsc{mis}}=p\, b^{N-13}$) for linear sweeps at different ramp rates. At intermediate ramping rates, we observe good agreement between the fits of the experimental and the simulated data. 
    For slower ramp rates (gray markers), the numerical data follows a superexponential trend because the dynamics resolve that the adiabatic gap decays superexponentially with the system size. This is shown in \textbf{(c)}, where we plot the simulated MIS probability as a function of the atom number. An exponential fit of the numerical data is not meaningful in this fast regime. Experimentally, however, the data still follows an exponential trend due to small hardware imperfections (cf.~Fig.~\ref{fig:Fig2}). The data highlights that the sweep-quench-sweep protocol [black dashed line in  \textbf{(a)}] outperforms linear sweeps over the accessible ramp rates.}
    \label{fig:AppSweepRates}
\end{figure}

Experimentally, we have observed that the scaling behavior is markedly different between a slow and a fast sweep. Here, we provide additional experimental data and also numerical tensor-network simulations to explain this behavior. We consider a linear sweep for different system sizes $N$ with various sweep rates, and find two different regimes (Fig.~\ref{fig:AppSweepRates}). This study is motivated by the question of whether the improvement yielded by the SQS protocol is reliant on the coherent dynamics during the duration of the quench or whether any generic nonadiabatic process facilitating population transfer to excited states would suffice. To investigate this, we fit the obtained MIS probabilities as $P_{\textsc{mis}}$\,$=$\,$p\, b^{N-13}$ by fitting a line to $\log(P_{\textsc{mis}})$. We observe in Fig.~\ref{fig:Fig2}(g), that for smaller system sizes, the slow sweep outperforms the fast one (since the gap $g_{\text{min}}$ is large enough to allow maintaining adiabaticity) but for sufficiently long chains, the latter yields a higher $P_{\textsc{mis}}$. However, examining the base of the fit $b$, we find that the SQS protocol demonstrates a better scaling than both slow and fast ramps, as well as \textit{all} ramps with intermediate sweep rates [Fig.~\ref{fig:AppSweepRates}(a)].

Interestingly, although one cannot directly establish the theoretically predicted superexponential behavior given the range of available system sizes, for slow sweeps, we indeed find a large prefactor $p$ in Fig.~\ref{fig:AppSweepRates}(b), which hints at such a scaling. For faster sweeps, the gap cannot be resolved in the sense that the size of the small gap does not influence the probability anymore. Instead, the population is diabatically transferred to higher excited states to a small extent and then tunnels through the avoided crossing. The success probability of this process is exponentially small. 

We conduct numerical simulations to confirm these observations in Fig.~\ref{fig:AppSweepRates}. Starting from the ground state at $\Delta_i=-4\Omega$, we sweep up to a final value of $\Delta_f=4\Omega$, as in the experiment. The simulation includes, as before, interactions up to next-nearest neighbors and considers a spacing of $s = 5.5$\;\textmu m between nearest neighbors. For relatively slow sweeps, we observe a scaling of the MIS probability in Fig.~\ref{fig:AppSweepRates}(c) which is consistent with a superexponential scaling of the adiabatic gap. In fact, there is an initial plateau and then the curves roll off very steeply. This plateau can be explained by the number of doublets in the Rydberg doublet chain. For a system of, say, $N=16$ atoms, the chain length is $L=11$, and there are five doublets in total. Due to boundary effects, the outermost doublets do not participate in the mechanism for the superexponential gap, which is why there are effectively only three doublets that contribute to the gap scaling. Therefore, the superexponential gap scaling is not pronounced for smaller system sizes.
When the sweep rate is faster, the data suggests that some exponentially small fraction of the population is first transferred to a low-energy eigenstate and then continues through the avoided level crossing to the MIS state. 

\clearpage
\bibliography{biblio}

\end{document}